\documentclass[11pt]{article}
\usepackage{epsfig}
\author{S. Sivasubramanian, Y.N. Srivastava$^\dagger $
G. Vitiello$^{\dagger\dagger }$\footnote{communicating author:
vitiello@sa.infn.it}~, A. Widom\\ \\
Physics Department, Northeastern University, Boston, MA USA \\
$^{\dagger}$Dipartimento di Fisica \& INFN, Universit\'a di Perugia,
Perugia, Italia\\
$^{\dagger \dagger }$Dipartimento di Fisica ``E.R. Caianiello'',
INFN \& INFM,\\ Universit\'a degli Studi di Salerno, Salerno, Italia}

\title{Quantum Dissipation Induced\\ Noncommutative Geometry}
\date{}

\begin{document}
\maketitle

\vskip .5cm
\centerline{Abstract}
\medskip
\par \noindent
{The quantum statistical dynamics of a position coordinate {\it x}
coupled to a reservoir requires theoretically two copies of the
position coordinate within the reduced density matrix description.
One coordinate moves forward in time while the other coordinate
moves backward in time. It is shown that quantum dissipation
induces, in the plane of the forward and backward motions, a
noncommutative geometry. The noncommutative geometric plane is a
consequence of a quantum dissipation induced phase interference
which is closely analogous to the Aharanov-Bohm effect.}

geometry, gauge theories, dissipation


\section{Introduction}

There has been considerable recent interest in the role of
noncommutative geometry in quantum mechanics. The algebraic
structures arising in that context have been
analyzed\cite{nc1}-\cite{Banerjee:2001zi}. In his work on the energy
levels of a charged particle in a magnetic field, Landau
pointed out the non-commuting nature of the coordinates of the
center of the circular cyclotron trajectory\cite{Landau}. The
harmonic oscillator on the noncommutative plane, the motion of a
particle in an external magnetic field and the Landau problem on
the noncommutative sphere are only few examples of systems which
have been studied in detail. Furthermore, noncommutative
geometries are also of interest in Cern-Simons gauge theories,
the usual gauge theories and string theories\cite{Dunne:1989hv}
-\cite{Banerjee:2001un}. Non-zero Poisson brackets for the plane
coordinates have been found in the study of
the symplectic structure for a dissipative system in the case of
strong damping $R \gg M$, i.e. the so-called reduced
case\cite{Blasone:1996yh}. The relation between dissipation
and noncommutative geometry was also noticed\cite{Banerjee:2001zi}
with reference to the reduced dissipative systems. The purpose of
the present paper is twofold: (i) we show
that quantum dissipation introduces its own noncommutative
geometry and (ii) we prove that the quantum interference phase
between two alternative paths in the plane (as in the
Aharanov-Bohm effect) is simply determined by the noncommutative
length scale and the enclosed area between the paths. This in turn
provides the connection between the noncommutative length scale
and the zero point fluctuations in the coordinates. The links we
establish between noncommutative geometry, quantum dissipation,
geometric phases and zero point fluctuations, may open interesting
perspectives in many sectors of quantum physics, e.g. in quantum
optics and in quantum computing or whenever quantum dissipation
cannot be actually neglected in any reasonable approximation. They
may also play a role in the 't Hooft proposal\cite{'tHooft:1999gk}
of the interplay between classical
deterministic systems with loss of information and quantum
dynamics in view of the established relation in that frame between
geometric phase and zero point energy\cite{Blasone:2000ew}.

Perhaps the clearest example of a noncommutive geometry is the
``plane''. Suppose that \begin{math} (X,Y)  \end{math} represents
the coordinates of a ``point'' in such a plane and further suppose
that the coordinates do not commute; i.e.
\begin{equation}
\left[X,Y\right]=iL^2,
\label{qauntcom1}
\end{equation}
where \begin{math} L \end{math} is the geometric length scale
in the noncommutative plane. The physical meaning of
\begin{math} L  \end{math} becomes evident upon
placing
\begin{equation}
Z=\frac{X+iY}{L\sqrt{2}},
\ \ Z^*=\frac{X-iY}{L\sqrt{2}},
\label{quantcom2}
\end{equation}
and
\begin{equation}
\left[Z,Z^* \right]=1,
\label{qauntcom3}
\end{equation}
into the noncommutative Pythagoras' definition of distance
\begin{math} S \end{math}; It is
\begin{equation}
S^2=X^2+Y^2=L^2(2Z^* Z+1).
\label{quantcom4}
\end{equation}
From the known properties of the oscillator destruction
\begin{math} Z \end{math} and creation \begin{math} Z^* \end{math}
operators in Eqs.(\ref{qauntcom3}) and (\ref{quantcom4}),
it follows that the Pythagorean distance is
quantized in units of the length scale \begin{math} L  \end{math}
according to
\begin{equation}
S_n^2=L^2(2n+1)\ \ {\rm where}\ \ n=0,1,2,3,\ldots \ .
\label{quantcom5}
\end{equation}

A quantum interference phase of the Aharanov-Bohm type can always
be associated with the noncommutative plane. From a path integral quantum
mechanical viewpoint, suppose that a particle can move from an initial point
in the plane to a final point in the plane via one of two paths, say
\begin{math} {\cal P}_1  \end{math} or \begin{math} {\cal P}_2 \end{math}.
Since the paths start and finish at the same point, if one
transverses the first path in a forward direction and the second
path in a backward direction, then the resulting closed path
encloses an area \begin{math} {\cal A} \end{math}. The phase
interference between these two points is determined by the
difference between the actions for these two paths
\begin{math} \hbar \vartheta ={\cal S}({\cal P}_1)-{\cal S}({\cal P}_2) \end{math}.
It will be shown in Sec.2 that the interference phase may be written as
\begin{equation}
\vartheta = \frac{\cal A}{L^2}.
\label{IntPhase}
\end{equation}

A physical realization of the mathematical noncommutative plane is present
in every laboratory wherein a charged particle moves in a plane with a normal
uniform magnetic field \begin{math} {\bf B} \end{math}. The nature of the
noncommutative geometry and the Aharanov-Bohm effect which follows from
the more general Eq.(\ref{IntPhase}) will be discussed in Sec.3.
For this case, there are two
canonical pairs of position coordinates which do not commute;
Namely, (i) the position \begin{math} {\bf R} \end{math} of the center of the
cyclotron circular orbit and (ii) the radius vector \begin{math} {\bf \rho } \end{math}
from the center of the circle to the charged particle position
\begin{math} {\bf r}={\bf R}+{\bf \rho } \end{math}.
This is shown in Fig.1. The magnetic length scale of the noncommuting geometric
coordinates is due to Landau,
\begin{equation}
L^2=\frac{\hbar c}{eB}=\frac{\phi_0}{2\pi B}
\ \ \ {\rm (magnetic)},
\label{MagneticLength}
\end{equation}
where \begin{math} \phi_0 \end{math} is the magnitude of the magnetic flux
quantum associated with a charge \begin{math} e \end{math}.

\begin{figure}[htbp]
\begin{center}
\mbox{\epsfig{file=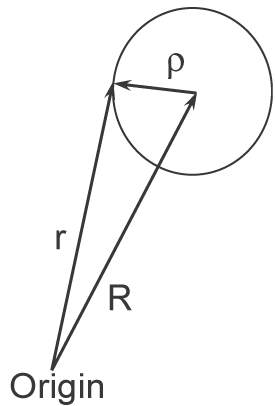,width=3.5cm}} \caption{\it Shown is
a charge $e$ moving in a circular cyclotron orbit. A noncommuting
coordinate pair is ${\bf R}=(X,Y)$ which points from the origin to
the orbit center. Another noncommuting coordinate pair is ${\bf
\rho }=(\rho_x,\rho_y)$ which points from the center of the orbit
to the charge position ${\bf r}={\bf R}+{\bf \rho}$.}
\label{ncfig1}
\end{center}
\end{figure}

Our main purpose here is to show that quantum dissipation introduces its
own noncommutative geometry. The dissipative noncommutative plane consists
of coordinates \begin{math} (x_+,x_-) \end{math} wherein \begin{math} x_+ \end{math}
denotes the coordinate \begin{math} x \end{math} moving forward in time
and \begin{math} x_- \end{math} denotes the coordinate
\begin{math} x \end{math} moving backward in time. The need for
``two copies'' \begin{math} (x_+,x_-) \end{math} for one physical
coordinate \begin{math} x \end{math} is a consequence of employing a
reduced density matrix
\begin{math} \left(x_+\left|\rho (t) \right|x_-\right) \end{math} for
describing quantum mechanical probabilities
\cite{Feynman}-\cite{Blasone:1997xt}.

The reader is asked to consider a particle moving in a potential
\begin{math} U(x) \end{math} under the further action of a
linear force of friction. Classically, the equation of motion would be
\begin{equation}
M\frac{d^2x}{dt^2}+R\frac{dx}{dt}+U^\prime (x)=0.
\label{quantcom6}
\end{equation}
In the ``two coordinate'' quantum mechanical version,
Eq.(\ref{quantcom6}) reads
\begin{eqnarray}
M\frac{d^2x_+}{dt^2}+R\frac{dx_-}{dt}+U^\prime (x_+)&=&0,
\nonumber \\
M\frac{d^2x_-}{dt^2}+R\frac{dx_+}{dt}+U^\prime (x_-)&=&0.
\label{quantcom7}
\end{eqnarray}
The dissipative system will behave classically,
as in Eq.(\ref{quantcom6}), if and only if
the forward and backward paths are nearly equal
\begin{math} x_+(t)\approx x_-(t)\approx x(t)\end{math}.
On the other hand, if \begin{math} x_+(t)\end{math} is appreciably
different from \begin{math} x_-(t)\end{math}, then quantum
interference will occur albeit in the presence of quantum
dissipation \cite{Blasone:1997xt}. It is remarkable the quantum
interference can in fact be induced by dissipation. The derivation
of Eqs.(\ref{quantcom7}) will be discussed in Sec.4.

In Sec.5, the noncommuting position coordinates are introduced in
the \begin{math} (x_+,x_-) \end{math} plane. The dissipation
induced length scale \begin{math} L \end{math} is determined by
\begin{equation}
L^2=\frac{\hbar}{R}
\ \ \ {\rm (dissipative)}.
\label{DissipationLength}
\end{equation}
Attention will then be paid to the situation in which the potential
\begin{math} U=0 \end{math} so that the force on the particle is {\em purely}
that of friction. The situation is then closely analogous to
the charged particle in an electric field since
the cyclotron circular orbits here appear as Minkowski metric hyperbolic orbits.
In the concluding Sec.6, the general physical basis for the quantum dissipation
induced noncommuting geometry will be discussed.

\section{Path Integrals and the Interference Phase}
For motion at fixed energy one may (in classical mechanics) associate with each path
\begin{math} {\cal P} \end{math} (in phase space) a phase space action integral
\begin{equation}
{\cal S}({\cal P})=\int_{\cal P} p_i dq^i.
\label{phase1}
\end{equation}
From the viewpoint of the path integral formulation of quantum mechanics one
may consider many possible paths with the same initial point and final point.
Let us concentrate on just two such paths
\begin{math} {\cal P}_1 \end{math} and \begin{math} {\cal P}_2 \end{math}.
The phase interference \begin{math} \vartheta  \end{math} between the two paths
is determined by the action difference
\begin{equation}
\hbar \vartheta =\int_{{\cal P}_1} p_i dq^i-\int_{{\cal P}_2} p_i dq^i
=\oint_{{\cal P}=\partial \Omega } p_i dq^i
\label{phase2}
\end{equation}
wherein \begin{math} {\cal P} \end{math} is the closed path which goes
from the initial point to the final point via path
\begin{math} {\cal P}_1 \end{math} and returns back to the initial point
via \begin{math} {\cal P}_2 \end{math}. The closed
\begin{math} {\cal P} \end{math} path may be regarded
as the boundary of a two-dimensional surface
\begin{math} \Omega \end{math}; i.e.
\begin{math} {\cal P}=\partial \Omega \end{math}.
Employing Stokes theorem in Eq.(\ref{phase2}) yields
\begin{equation}
\vartheta = \frac{1}{\hbar }\oint_{{\cal P}=\partial \Omega } p_i dq^i
=\frac{1}{\hbar }\int _\Omega (dp_i \wedge dq^i).
\label{phase3}
\end{equation}
The quantum phase interference \begin{math} \vartheta  \end{math}
between two alternative paths is thereby proportional to an ``area''
of a surface \begin{math} \Omega  \end{math}
in phase space \begin{math} (p_1,\ldots ,p_f;q^1,\ldots ,q^f) \end{math}
as described by the right hand side of Eq.(\ref{phase3}).

If one briefly reverts to the operator formalism and writes
the commutation Eq.(\ref{qauntcom1}) in the noncommutative
plane as
\begin{equation}
[X,P_X]=i\hbar \ \ \ {\rm where}
\ \ \ P_X=\left(\frac{\hbar Y}{L^2}\right),
\label{phase4}
\end{equation}
then back in the path integral formalism Eq.(\ref{phase3}) reads
\begin{equation}
\vartheta =\frac{1}{\hbar }\int _\Omega (dP_X \wedge dX)
=\frac{1}{L^2}\int _\Omega (dY \wedge dX)
\end{equation}
and we have proved the following:
\par \noindent
{\bf Theorem:} {\em The quantum interference phase between two alternative paths
in the plane is determined by the noncommutative length scale
\begin{math} L  \end{math} and the enclosed area
\begin{math} {\cal A} \end{math} via
\begin{equation}
\vartheta = \frac{\cal A}{L^2}\ .
\label{IntPhaseTheorem}
\end{equation}}
\par \noindent
The existence of an interference phase is intimately connected to the
zero point fluctuations in the coordinates; e.g. Eq.(\ref{qauntcom1})
implies a zero point uncertainty relation
\begin{math} \Delta X \Delta Y \ge (L^2/2) \end{math}.

\section{Charged Particle in a Magnetic Field}
Consider the motion of a non-relativistic charged particle in a plane
perpendicular to a uniform magnetic field \begin{math} {\bf B} \end{math}.
The Hamiltonian is
\begin{equation}
H=\frac{1}{2M}\left({\bf p}-\frac{e}{c}{\bf A}\right)^2
\ \ {\rm where}\ \ {\bf A}=(A_x,A_y)=\left(-\frac{By}{2},\frac{Bx}{2}\right).
\label{magnetic1}
\end{equation}
Putting
\begin{equation}
\hbar {\bf K}={\bf p}-\frac{e}{c}{\bf A}=-i\hbar {\bf \nabla}-\frac{e}{c}{\bf A},
\label{magnetic2}
\end{equation}
yields the equal time commutator
\begin{equation}
\left[K_x,K_y\right]=\frac{ieB}{\hbar c}=\frac{i}{L^2}.
\label{magnetic3}
\end{equation}
Let us now define the cyclotron radius vector
\begin{math} {\bf \rho}=(\rho_x,\rho_y) \end{math} as
\begin{equation}
\rho_x=-L^2K_y\ \ {\rm and}\ \ \rho_y=L^2K_x.
\label{magnetic4}
\end{equation}
If \begin{math} M{\bf v}=\hbar {\bf K} \end{math} were classical,
then \begin{math} {\bf \rho } \end{math} would be the radius vector
from the center of the circular cyclotron orbit to the position of the
charge. When quantum mechanics is employed, the notion of a cyclotron
orbit becomes blurred because the vector cyclotron radius has components
which are noncommutative,
\begin{equation}
\left[\rho_x,\rho_y\right]=iL^2.
\label{magnetic5}
\end{equation}
The energy of the charged particle may still be written in terms of the
cyclotron radius \begin{math} {\bf \rho }  \end{math} and the cyclotron
frequency \begin{math} \omega_c  \end{math} as
\begin{equation}
H=\frac{1}{2}M\omega_c^2\rho^2
=\frac{1}{2}M\omega_c^2(\rho_x^2+\rho_y^2)
\ \ {\rm where}
\ \ \omega_c=\frac{eB}{Mc}\ .
\label{magnetic6}
\end{equation}
The noncommutative geometrical Pythagorean theorem yields
the quantized radius vector values
\begin{equation}
\rho_n^2=L^2(2n+1)\ \ \ \ (n=0,1,2,3,\ldots ).
\label{magnetic7}
\end{equation}
Eqs.(\ref{magnetic6}) and (\ref{magnetic7}) imply the
Landau magnetic energy spectrum
\begin{equation}
E_n=\hbar \omega_c \left(n+\frac{1}{2}\right)\ \ \ \ (n=0,1,2,3,\ldots ).
\label{magnetic8}
\end{equation}

Note that the position of the charge \begin{math} {\bf r}=(x,y) \end{math}
has components which commute \begin{math} [x,y]=0 \end{math}, but these do
not commute with the cyclotron radius components; i.e. we have from
Eqs.(\ref{magnetic2}) and ({\ref{magnetic4}}) that
\begin{eqnarray}
\left[\rho_x,x\right]&=&\left[\rho_y,y\right]\ =\ 0, \nonumber \\
\left[\rho_x,y\right]&=&\left[x,\rho_y \right]\ =\ iL^2.
\label{magnetic9}
\end{eqnarray}
We then introduce the coordinate
\begin{math} {\bf R}=(X,Y) \end{math}
as the center of the cyclotron orbit via
\begin{equation}
{\bf r}={\bf R}+{\bf \rho }
\label{magnetic10}
\end{equation}
and find that
\begin{equation}
\left[X,Y\right]=-iL^2.
\label{magnetic11}
\end{equation}
Thus \begin{math} {\bf R}=(X,Y) \end{math} and
\begin{math} {\bf \rho}=(\rho_x,\rho_y) \end{math}
represent two independent pairs of geometric
canonical conjugate variables; i.e
\begin{math} [R_i,\rho_j]=0 \end{math}.

From a path integral viewpoint, the quantum interference
phase in the plane is described by Eq.(\ref{IntPhaseTheorem}).
For the magnetic field problem the theorem reads
\begin{equation}
\vartheta =\frac{\cal A}{L^2}=
\frac{\cal A}{\hbar c/eB}=\frac{e\Phi }{\hbar c},
\label{magnetic12}
\end{equation}
where \begin{math} \Phi=B{\cal A} \end{math} is the magnetic flux
through the enclosed area \begin{math} {\cal A} \end{math}.
Eq.(\ref{magnetic12}) represents precisely the Aharanov-Bohm effect.

Finally, in the operator (as opposed to path integral) version of
quantum mechanics, the area enclosed by the cyclotron orbit in the
plane has the discrete spectrum
\begin{equation}
{\cal A}_n=\pi \rho_n^2 =\pi L^2 (2n+1)
=\frac{2\pi \hbar c}{eB}\left(n+\frac{1}{2}\right).
\label{magnetic13}
\end{equation}
This means that as the radii of the cyclotron increase, the added
magnetic flux comes in units of the flux quantum
\begin{math} \phi_0 \end{math}; i.e.
\begin{equation}
\Delta \Phi =B({\cal A}_{n+1}-{\cal A}_n)
=\frac{2\pi \hbar c}{e}=\phi_0.
\label{magnetic14}
\end{equation}
Such magnetic flux quantization here arises as a consequence of the area
quantization which is intrinsic to the noncommutative plane.

\section{Quantum Friction}
The quantum properties of a ``position coordinate'' \begin{math} x \end{math}
of a particle are best described by making ``two copies'' of the coordinate
\begin{math} x\to (x_+,x_-) \end{math} \cite{Feynman}-\cite{Blasone:1997xt}.
For computing averages of any possible
associated operator (say \begin{math} Q \end{math}) employing a reduced
density matrix (say \begin{math} \rho \end{math}) one must integrate over both
coordinates (\begin{math} x_+ \end{math} and \begin{math} x_- \end{math}) in
the copies; i.e. the averaged value of \begin{math} Q \end{math} is of the form
\begin{equation}
\left<Q \right>=Tr\left(\rho Q\right)
=\int \int \left( x_+ \left| \rho \right| x_- \right)
\left( x_- \left| Q \right| x_+ \right) dx_+ dx_-.
\label{QD1}
\end{equation}
For a particle moving in one dimension with a Hamiltonian
\begin{equation}
H=\frac{p^2}{2M}+U(x)=
-\frac{\hbar^2}{2M}\left(\frac{\partial }{\partial x}\right)^2+U(x),
\label{QD2}
\end{equation}
the time dependence,
\begin{equation}
\rho(t)=e^{-iHt/\hbar }\rho e^{iHt/\hbar },
\label{QD3}
\end{equation}
reads (in the coordinate representation)
\begin{equation}
\left( x_+ \left| \rho (t) \right| x_- \right)=
e^{-i(H_+-H_-)t/\hbar }\left( x_+ \left| \rho \right| x_- \right),
\label{QD4}
\end{equation}
where the two copies of the Hamiltonian
\begin{equation}
H_\pm=\frac{p_\pm^2}{2M}+U(x_\pm)=
-\frac{\hbar^2}{2M}\left(\frac{\partial}{\partial x_\pm}\right)^2+U(x_\pm )
\label{QD5}
\end{equation}
drive \begin{math} x_+ \end{math} forward in time and drive
\begin{math} x_-  \end{math} backward in time.
The equation of motion for the density matrix is then
\begin{equation}
i\hbar \frac{\partial }{\partial t}\left( x_+ \left| \rho (t) \right| x_- \right)
={\cal H}_0\left( x_+ \left| \rho (t) \right| x_- \right)
\label{QD6}
\end{equation}
wherein
\begin{equation}
{\cal H}_0=H_+ - H_-=
\frac{p_+^2}{2M}+U(x_+)-\frac{p_-^2}{2M}-U(x_-).
\label{QD7}
\end{equation}
The notion of quantum dissipation enters into our considerations
if there is a coupling to a thermal reservoir yielding
a mechanical resistance \begin{math} R \end{math}. The full equation
of motion has the form
\begin{equation}
i\hbar \frac{\partial }{\partial t}\left( x_+ \left| \rho (t) \right| x_- \right)
={\cal H}\left( x_+ \left| \rho (t) \right| x_- \right)
-\left( x_+ \left|N[\rho ] \right| x_- \right),
\label{QD8}
\end{equation}
where \begin{math} N[\rho ]\approx (ik_BTR/\hbar )[x,[x,\rho ]]
\end{math} describes the effects of the reservoir random thermal
noise and the new ``Hamiltonian'' \begin{math} {\cal H}_0\to {\cal
H}  \end{math} for motion in the \begin{math} (x_+,x_-)
\end{math} plane has been previously
discussed\cite{Srivastava:1995yf,Blasone:1997xt}
\begin{equation}
{\cal H}=\frac{1}{2M}\left\{\left(p_+ -\frac{Rx_-}{2}\right)^2
-\left(p_-+ \frac{Rx_+}{2}\right)^2 \right\}+U(x_+)-U(x_-).
\label{QD9}
\end{equation}

The velocity components \begin{math} (v_+,v_-) \end{math} in the
\begin{math} (x_+,x_-) \end{math} plane may be found from
the Hamiltonian equation
\begin{equation}
v_\pm =\dot{x}_\pm = \frac{\partial {\cal H}}{\partial p_\pm }=
\pm \frac{1}{M}\left(p_\pm \mp \frac{Rx_\mp }{2} \right).
\label{QD10}
\end{equation}
Similarly,
\begin{equation}
\dot{p}_\pm =- \frac{\partial {\cal H}}{\partial x_\pm }
=\mp U^\prime (x_\pm )\mp \frac{Rv_\mp }{2}\ .
\label{QD11}
\end{equation}
From Eqs.(\ref{QD10}) and (\ref{QD11}) it follows that
\begin{equation}
M\dot{v}_\pm +Rv_{\mp}+U^\prime (x_\pm )=0
\label{QD12}
\end{equation}
in agreement with Eqs.(\ref{quantcom7}).
The classical equation of motion including dissipation
thereby holds true if
\begin{math} x_+(t)\approx x_-(t)\approx x(t)\end{math}.
Dissipation induced quantum interference takes place if and only
if the forward in time paths differ appreciably from the backward
in time paths\cite{Blasone:1997xt}.

\section{Dissipative Noncommutative Plane}
The commutation relations in the dissipative
\begin{math} (x_+,x_-) \end{math} plane may now be derived.
If we define
\begin{equation}
Mv_\pm =\hbar K_\pm ,
\label{momentum}
\end{equation}
then one finds from Eq.(\ref{QD10}) that
\begin{equation}
\left[K_+,K_-\right]=\frac{iR}{\hbar }
=\frac{i}{L^2}\ .
\label{DP1}
\end{equation}
A canonical set of conjugate position coordinates
\begin{math} (\xi_+,\xi_-) \end{math} may be defined by
\begin{eqnarray}
\xi_\pm &=& \mp L^2K_\mp \nonumber \\
\left[\xi_+,\xi_-\right] &=& iL^2.
\label{DP2}
\end{eqnarray}
Another canonical set of conjugate position coordinates
\begin{math} (X_+,X_-) \end{math} may be defined by
\begin{eqnarray}
x_+=X_+ +\xi_+ &,& x_-=X_- +\xi_-
\nonumber \\
\left[X_+,X_-\right] &=& -iL^2.
\label{DP3}
\end{eqnarray}
Note that \begin{math} [X_a,\xi_b]=0 \end{math} , where
\begin{math} a=\pm \end{math} and  \begin{math} b=\pm \end{math}.

\begin{figure}[htbp]
\begin{center}
\mbox{\epsfig{file=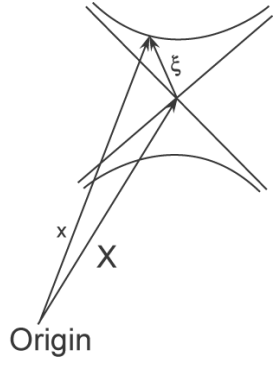,width=3.5cm}} \caption{\it Shown is
the hyperbolic path of a particle moving in the $x=(x_+,x_-)$
plane. A noncommuting coordinate pair is $X=(X_+,X_-)$ which
points from the origin to hyperbolic center. Another noncommuting
coordinate pair is $\xi =(\xi_+, \xi_- )$ which points from the
center of the orbit to the position on the hyperbola. $x=X+\xi$.}
\label{ncfig2}
\end{center}
\end{figure}

For the case of pure friction in which the potential
\begin{math} U=0 \end{math}, Eqs.(\ref{QD9}),
(\ref{momentum}) and (\ref{DP2}) imply
\begin{equation}
{\cal H}_{friction}=\frac{\hbar^2}{2M}(K_+^2-K_-^2)
=-\frac{\hbar^2}{2ML^4}(\xi_+^2-\xi_-^2).
\label{DP4}
\end{equation}
The equations of motion read
\begin{equation}
\dot{\xi}_\pm =\frac{i}{\hbar}\left[{\cal H}_{friction},\xi_\pm\right]
=-\frac{\hbar}{ML^2}\xi_\mp
=-\frac{R}{M}\xi_\mp =-\Gamma \xi_\mp ,
\label{DP5}
\end{equation}
with the solution
\begin{equation}
\left(\begin{array}{c}
  \xi_+(t)\\
  \xi_-(t) \\
\end{array}\right)=
\left(\begin{array}{cc}
  \cosh(\Gamma t) & -\sinh(\Gamma t) \\
  -\sinh(\Gamma t) & \cosh(\Gamma t) \\
\end{array}\right)
\left(\begin{array}{c}
  \xi_+ \\
  \xi_- \\
\end{array}\right).
\label{DP6}
\end{equation}
Eq.(\ref{DP6}) describes the hyperbolic orbit
\begin{equation}
\xi_-(t)^2-\xi_+(t)^2=
\frac{2L^2{\cal{H}}_{friction}}{\hbar \Gamma }\ .
\label{DP7}
\end{equation}

A comparison can be made between the noncommutative dissipative plane
and the noncommutative Landau magnetic plane
as shown in Fig.2. The circular orbit in Fig.1 for the
magnetic problem is here replaced by the hyperbolic orbit.
In view of the minus sign in the ``kinetic'' energy,
\begin{equation}
{\cal H}_{friction}=\frac{M}{2}(v_+^2-v_-^2),
\label{DP8}
\end{equation}
it is best to view the metric as pseudo-Euclidean or
equivalently we can use the Minkowski metric
\begin{math}
u\cdot w=u_+w_+ - u_-w_-
\end{math} .

In fact, the quantum dissipative eigenvalue problem
\begin{math}
{\cal H}_{friction}\ \tilde{\rho}_\omega =
\hbar \omega \tilde{\rho}_\omega
\end{math}
is formally identical to the relativistic charged scalar
field equation in \begin{math} (1+1) \end{math} dimensional
quantum electrodynamics; i.e.
$$
\left\{-d_\mu d^\mu +
\left(\frac{mc}{\hbar}\right)^2\right\}\psi (x) = 0,
$$
$$
K_\mu =-id_\mu = -i\partial_\mu + \frac{eA_\mu}{\hbar c},
$$
\begin{equation}
\left[K_\mu ,K_\nu \right] = \frac{i\hbar eF_{\mu \nu }}{c}\ .
\label{QED1}
\end{equation}
Since in \begin{math} (1+1) \end{math} dimensional
electrodynamics, the only nonzero tensor
components describe the electric field
\begin{math} F_{10}=-F_{01}=E \end{math},
it follows by comparing Eqs.(\ref{DP1}) and (\ref{QED1})
that the analogy is exact if
\begin{math} L^2=(\hbar/R)=(\hbar c/eE) \end{math}. Note
that the interference phase is thereby
\begin{eqnarray}
\vartheta &=& \frac{e\Phi }{\hbar c}
=\frac{e}{\hbar c}\oint_{{\cal P}\partial \Omega } A_\nu dx^\nu
\nonumber \\
&=& \frac{e}{2\hbar c}\int_\Omega F_{\mu \nu} dx^\mu \wedge dx^\nu
\nonumber \\
&=& \frac{eE{\cal A}}{\hbar c}=\frac{\cal A}{L^2}\ .
\label{QED2}
\end{eqnarray}
Thus, the Minkowski metric implies a closer analogy with the
electric flux than with the magnetic flux. The hyperbolic
orbit in Fig.2 is reflected in the classical orbit for a charged particle
moving along the \begin{math} x \end{math}-axis in a uniform electric
field. The hyperbolae are defined by
\begin{math}
(x-X)^2-c^2(t-T)=\Lambda^2 ,
\label{PairProduct}
\end{math}
where \begin{math} \Lambda^2=(mc^2/eE)^2=(mc/\hbar L^2)^2  \end{math},
the hyperbolic center is at \begin{math} (X,cT) \end{math} and
one branch of the hyperbolae is a charged particle moving forward
in time while the other branch is the same particle moving backward
in time as an anti-particle.

\section{Remarks and Conclusions}

We have discussed above the dissipative quantum statistical dynamics
of a coordinate \begin{math} x \end{math} coupled
to a reservoir yielding friction effects. The
reduced density matrix description requires theoretically two copies of the
position coordinate \begin{math} (x_+,x_-) \end{math} with \begin{math} x_+ \end{math}
moving forward and \begin{math} x_- \end{math} moving backward in time.
Both decay \begin{math} \exp(-\Gamma t)  \end{math} and amplification
\begin{math} \exp(\Gamma t)  \end{math} enter into the motions of
\begin{math} x_\pm (t) \end{math}. Quasi-classically, the motions proceed
on two branches of hyperbolae whose center \begin{math} X=(X_+,X_-) \end{math}
and relative displacements \begin{math} \xi=(\xi_+,\xi_-) \end{math} are
independent canonical conjugate pairs which obey the rules of the non-commutative
plane; i.e.
\begin{eqnarray}
\left[\xi_+,\xi_- \right] &=& iL^2 \nonumber \\
\left[X_+,X_- \right] &=& -iL^2
\label{noncom}
\end{eqnarray}
The noncommutative geometric plane is intimately related to an interference phase
which can serve as a quantum basis for deriving the canonical commutation relations
between coordinates in the plane. For completeness of presentation we derive in the
appendix to this work the noncommuting geometry vortex coordinates\cite{mittag}
in thin superfluid films possibly induced by rotations of the superfluid container or
substrate\cite{Widom}.

For the problem at hand, a density matrix equation of motion
\begin{math} i\hbar \dot{\rho}={\cal H}\rho \end{math}
leads to an eigenvalue problem
\begin{math} {\cal H}\rho_\omega =\hbar \omega \rho_\omega \end{math}
which directly yields the quantum frequency spectrum. If the system
were isolated, then \begin{math} {\cal H}=H_+ - H_- \end{math}.
For isolated quantum systems, the frequencies can be identified with
the Bohr transition frequencies \begin{math} \omega_{fi}=E_f-E_i \end{math}.
Thus a quantum jump \begin{math} i\to f \end{math} involves a transition
from a backward in time motion to a forward in time motion as is evident
from
\begin{math}
\exp(-i\omega_{fi}t)=
\exp(-iE_ft/\hbar ) \exp(iE_it/\hbar ).
\end{math}

If the quantum system is not isolated,
then the forward in time to the backward in time transitions
are strongly coupled. Nevertheless the eigenvalue problem
\begin{equation}
{\cal H}\rho_\omega =\hbar \omega \rho_\omega
\label{C1}
\end{equation}
still describes the quantum spectroscopic frequency spectrum.
For the problem of pure frictional damping, we have from
Eqs.(\ref{DP2}) and (\ref{DP4}) that
\begin{equation}
{\cal H}_{friction}=\frac{\Pi_+^2}{2M}-\frac{M\Gamma^2 }{2}\xi_+^2,
\ \ {\rm where}\ \ {\Pi }_+ =\frac{\hbar \xi_-}{L^2}\ .
\label{C2}
\end{equation}
Since \begin{math} [\Pi_+,\xi_+]=-i\hbar  \end{math}
the operator Eq.(\ref{C2}) represents an ``inverted oscillator''.
The barrier transmission coefficient \begin{math} P(\omega ) \end{math}
for the inverted oscillator is well known\cite{Landau}; It is
\begin{equation}
P(\omega )=\frac{1}{1+e^{-2\pi \omega/\Gamma}}\ .
\label{C3}
\end{equation}
Thus there is a possibility of jumping from the forward direction in
time to the backward direction of time or vice versa. Such quantum
jumps are required for the Bohr frequencies
\begin{math} \omega \end{math} in an open (dissipative) system.

We observe that our conclusions may be extended to the
three-dimensional topological massive Chern-Simons gauge theory in
the infrared limit and to the Bloch electron in a solid. We recall
indeed that the Lagrangian for the system of Eqs.(\ref{quantcom7})
has been found \cite{Blasone:1996yh} to be the same as the
Lagrangian for three-dimensional topological massive Chern-Simons
gauge theory in the infrared limit. It is also the same as for a
Bloch electron in a solid which propagates along a lattice plane
with a hyperbolic energy surface\cite{Blasone:1996yh}. In the
Chern-Simons case we have $\theta_{CS} = R/M = (\hbar/M L^2)$,
with $\theta_{CS} $ the ``topological mass parameter''. In the
Bloch electron case, $(eB/\hbar c) =(1/L^2)$, with $B$ denoting
the $z$-component of the applied external magnetic field. In ref.
\cite{Blasone:1996yh} (see also \cite{Banerjee:2001zi}) it has
been considered the symplectic structure for the system of Eqs.
(\ref{quantcom7}) in the case of strong damping $R \gg M$ (the
so-called reduced case) in the Dirac constraint formalism as well
as in the Faddeev and Jackiw formalism \cite{Faddeev} and in both
formalisms a non-zero Poisson bracket for the ($x_+,~x_-$)
coordinates has been found.

\vskip .5cm \centerline{\bf Acknowledgements}
\bigskip
\par \noindent
The authors would like to thank INFN, INFM and ESF Program COSLAB
for partial support of this work.

\vskip .5cm
\centerline{\bf APPENDIX}
\bigskip
In thin superfluid \begin{math} He^4 \end{math} films on solid substrates,
vortices can exist. The coordinates (x,y) locating the core of a single vortex
do not commute with each other and thus determine a noncommutative geometry.
Let us see how this comes about.

The superfluid velocity field \begin{math} {\bf v}_s \end{math} rotating
around the vortex core is related to the phase of the wave function
via
$$
{\bf v}_s={\hbar \over m}{\bf grad}\ (\sigma \theta) ,
$$
where $\sigma =\pm 1$ depending on the orientation of the vortex.
For such a flow, the many body wave function has the form
$$
\Psi \approx \exp \big(i\sigma \sum_j\theta ({\bf r}_j)\big)\Psi_0
$$
wherein $\Psi_0$ is real. Because $\theta $ is a phase, we have for
an integral around the core
$$
\oint_{(around\ core)}d\theta =2\pi .
$$
Thus one has the circulation quantization
$$
m\oint_{(around\ core)}{\bf v}_s\cdot d{\bf r}=2\pi \sigma \hbar
\ \ {\rm where} \ \ {(\sigma=\pm 1)}.
$$

Now let us consider what happens if we move the core around a closed path
${\cal P}=\partial \Omega $ wherein the enclosed area contains
${\cal N}(\Omega )$ adsorbed atoms. Each atom whose position obeys
${\bf r}_j \in \Omega $ receives a change of $2\pi \sigma $ in the total
phase. Each atom which obeys ${\bf r}_j \notin \Omega $ receives no phase
change. Thus the total phase change which takes place as the core is brought
around a closed path is given by
$$
\vartheta =2\pi \sigma {\cal N}(\Omega ).
$$
If $n={\cal N}(\Omega )/{\cal A}(\Omega)$ denotes the number of atoms per
unit area adsorbed on the film, then
$$
\vartheta=2\pi \sigma {\cal N}=2\pi \sigma n {\cal A}.
$$
Comparing the above equation with our central theorem
Eq.(\ref{IntPhaseTheorem}), we find that $L^2=(2\pi n)^{-1}$ and that
the position of the vortex cores on the substrate obey
$$
\left[x,y\right]=\left(\frac{i}{2\pi n}\right)\sigma
\ \ {\rm where} \ \ {(\sigma=\pm 1)}.
$$
The noncommutative geometry length scale $L$ is of the order of
the vortex core size.
\vskip 1cm


\end{document}